\documentclass[superscriptaddress,aps,prc,showkeys,showpacs,onecolumn,10pt]{revtex4-1}

\usepackage{amsmath}
\usepackage{graphicx}
\usepackage{color}
\graphicspath{{fig/}}
\begin{document}
\title{ \bf \boldmath Measurement of the $\eta\to\pi^+\pi^-\pi^0$
Dalitz plot distribution}

\newcommand*{\IKPUU}{Division of Nuclear Physics, Department of Physics and 
 Astronomy, Uppsala University, Box 516, 75120 Uppsala, Sweden}
\newcommand*{\ASWarsN}{Department of Nuclear Physics, National Centre for 
 Nuclear Research, ul.\ Hoza~69, 00-681, Warsaw, Poland}
\newcommand*{\IPJ}{Institute of Physics, Jagiellonian University, ul.\ 
 Reymonta~4, 30-059 Krak\'{o}w, Poland}
\newcommand*{\PITue}{Physikalisches Institut, Eberhard--Karls--Universit\"at 
 T\"ubingen, Auf der Morgenstelle~14, 72076 T\"ubingen, Germany}
\newcommand*{\Kepler}{Kepler Center f\"ur Astro-- und Teilchenphysik, 
 Physikalisches Institut der Universit\"at T\"ubingen, Auf der 
 Morgenstelle~14, 72076 T\"ubingen, Germany}
\newcommand*{\MS}{Institut f\"ur Kernphysik, Westf\"alische 
 Wilhelms--Universit\"at M\"unster, Wilhelm--Klemm--Str.~9, 48149 M\"unster, 
 Germany}
\newcommand*{\ASWarsH}{High Energy Physics Department, National Centre for 
 Nuclear Research, ul.\ Hoza~69, 00-681, Warsaw, Poland}
\newcommand*{\IITB}{Department of Physics, Indian Institute of Technology 
 Bombay, Powai, Mumbai--400076, Maharashtra, India}
\newcommand*{\Budker}{Budker Institute of Nuclear Physics of SB RAS, 
 11~akademika Lavrentieva prospect, Novosibirsk, 630090, Russia}
\newcommand*{\Novosib}{Novosibirsk State University,  
 2~Pirogova Str., Novosibirsk, 630090, Russia}
\newcommand*{\IKPJ}{Institut f\"ur Kernphysik, Forschungszentrum J\"ulich, 
 52425 J\"ulich, Germany}
\newcommand*{\JCHP}{J\"ulich Center for Hadron Physics, Forschungszentrum 
 J\"ulich, 52425 J\"ulich, Germany}
\newcommand*{\Bochum}{Institut f\"ur Experimentalphysik I, Ruhr--Universit\"at 
 Bochum, Universit\"atsstr.~150, 44780 Bochum, Germany}
\newcommand*{\ZELJ}{Zentralinstitut f\"ur Engineering, Elektronik und 
 Analytik, Forschungszentrum J\"ulich, 52425 J\"ulich, Germany}
\newcommand*{\Erl}{Physikalisches Institut, 
 Friedrich--Alexander--Universit\"at Erlangen--N\"urnberg, 
 Erwin--Rommel-Str.~1, 91058 Erlangen, Germany}
\newcommand*{\ITEP}{Institute for Theoretical and Experimental Physics, State 
 Scientific Center of the Russian Federation, Bolshaya Cheremushkinskaya~25, 
 117218 Moscow, Russia}
\newcommand*{\Giess}{II.\ Physikalisches Institut, 
 Justus--Liebig--Universit\"at Gie{\ss}en, Heinrich--Buff--Ring~16, 35392 
 Giessen, Germany}
\newcommand*{\IITI}{Department of Physics, Indian Institute of Technology 
 Indore, Khandwa Road, Indore--452017, Madhya Pradesh, India}
\newcommand*{\HepGat}{High Energy Physics Division, Petersburg Nuclear Physics 
 Institute, Orlova Rosha~2, Gatchina, Leningrad district 188300, Russia}
\newcommand*{\HeJINR}{Veksler and Baldin Laboratory of High Energiy Physics, 
 Joint Institute for Nuclear Physics, Joliot--Curie~6, 141980 Dubna, Russia}
\newcommand*{\Katow}{August Che{\l}kowski Institute of Physics, University of 
 Silesia, Uniwersytecka~4, 40-007, Katowice, Poland}
\newcommand*{\IAS}{Institute for Advanced Simulation, Forschungszentrum 
 J\"ulich, 52425 J\"ulich, Germany}
\newcommand*{\HISKP}{Helmholtz--Institut f\"ur Strahlen-- und Kernphysik, 
 Rheinische Friedrich--Wilhelms--Universit\"at Bonn, Nu{\ss}allee~14--16, 
 53115 Bonn, Germany}
\newcommand*{\Bethe}{Bethe Center for Theoretical Physics, Rheinische 
 Friedrich--Wilhelms--Universit\"at Bonn, 53115 Bonn, Germany}
\newcommand*{\IFJ}{The Henryk Niewodnicza{\'n}ski Institute of Nuclear 
 Physics, Polish Academy of Sciences, 152~Radzikowskiego St, 31-342 
 Krak\'{o}w, Poland}
\newcommand*{\NuJINR}{Dzhelepov Laboratory of Nuclear Problems, Joint 
 Institute for Nuclear Physics, Joliot--Curie~6, 141980 Dubna, Russia}
\newcommand*{\KEK}{High Energy Accelerator Research Organisation KEK, Tsukuba, 
 Ibaraki 305--0801, Japan} 
\newcommand*{\IMPCAS}{Institute of Modern Physics, Chinese Academy of 
 Sciences, 509 Nanchang Rd., Lanzhou 730000, China}
\newcommand*{\ASLodz}{Department of Cosmic Ray Physics, National Centre for 
 Nuclear Research, ul.\ Uniwersytecka~5, 90--950 {\L}\'{o}d\'{z}, Poland}

\author{P.~Adlarson}    \affiliation{\IKPUU}
\author{W.~Augustyniak} \affiliation{\ASWarsN}
\author{W.~Bardan}      \affiliation{\IPJ}
\author{M.~Bashkanov}   \affiliation{\PITue}\affiliation{\Kepler}
\author{F.S.~Bergmann}  \affiliation{\MS}
\author{M.~Ber{\l}owski}\affiliation{\ASWarsH}
\author{H.~Bhatt}       \affiliation{\IITB}
\author{A.~Bondar}      \affiliation{\Budker}\affiliation{\Novosib}
\author{M.~B\"uscher}   \altaffiliation[present address: ]{\PGI;\,\Dussel}\affiliation{\IKPJ}\affiliation{\JCHP}
\author{H.~Cal\'{e}n}   \affiliation{\IKPUU}
% ...

\newcommand*{\PGI}{Peter Gr\"unberg Institut, PGI--6 Elektronische 
 Eigenschaften, Forschungszentrum J\"ulich, 52425 J\"ulich, Germany}
\newcommand*{\Dussel}{Institut f\"ur Laser-- und Plasmaphysik, Heinrich--Heine 
 Universit\"at D\"usseldorf, Universit\"atsstr.~1, 40225 Düsseldorf, Germany}
\newcommand*{\Bern}{Albert Einstein Center for Fundamental Physics, University 
 of Bern, Sidlerstrasse~5, 3012 Bern, Switzerland}
\newcommand*{\Aachen}{III.~Physikalisches Institut~B, Physikzentrum, 
 RWTH Aachen, 52056 Aachen, Germany}
\newcommand*{\Delhi}{Department of Physics and Astrophysics, University of 
Delhi, Delhi--110007, India}
\newcommand*{\Mainz}{Institut f\"ur Kernphysik, Johannes 
 Gutenberg--Universit\"at Mainz, Johann--Joachim--Becher Weg~45, 55128 Mainz, 
 Germany}
\newcommand*{\Sheff}{Department of Physics and Astronomy, University of 
 Sheffield, Hounsfield Road, Sheffield, S3 7RH, United Kingdom}

\collaboration{WASA-at-COSY Collaboration}\noaffiliation

\date{\today}

\begin{abstract}
   Dalitz  plot  distribution of the  $\eta\to\pi^+\pi^-\pi^0$  decay  is
   determined using a data sample  of $1.2\cdot 10^7$ $\eta$ mesons 
from $pd\to ^3\textrm{He}\eta$ reaction at 1 GeV collected by the
 WASA detector at COSY.
\end{abstract}


\maketitle
\section{Introduction}
The   amplitude    of   the   isospin    violating   decays   $\eta\to
\pi^+\pi^-\pi^0$ and  $\eta\to\pi^0\pi^0\pi^0$ is dominated  by a term
proportional to the light  quark mass difference $(m_d-m_u)$ since the
electromagnetic           contribution          is          suppressed
\cite{Sutherland:1966zz,Baur:1995gc,Ditsche:2008cq}.   This  makes the
decays    a   sensitive    probe   of    the   light    quark   masses
\cite{Leutwyler:1996qg}.   The  leading  term  for the  partial  decay
widths of the two decay modes is proportional 
to $Q^{-4}$ where $Q^{2}$ is defined as the   following
combination  of the light  quark masses \cite{Kaplan:1986ru}:
 \begin{equation}
Q^2=\frac{m_s^2-\hat{m}^2}{m_d^2-m_u^2},
\ \ \hat{m}=\frac{1}{2}(m_u+m_d).
\end{equation}
The  determination of  the  $Q$ parameter  requires  knowledge of  the
experimental value  of at  least one of  the $\eta\to\pi^+\pi^-\pi^0$,
$\eta\to\pi^0\pi^0\pi^0$  partial decay  widths and  the corresponding
proportionality factors.

Experimental  determination  of  the  partial  decay  widths  requires
knowledge of the  $\eta$ radiative width, $\Gamma_{\gamma\gamma}$, and
the               relative               branching              ratios
$BR(\eta\to\pi^0\pi^0\pi^0)/BR(\eta\to\gamma\gamma)$                and
$BR(\eta\to\pi^+\pi^-\pi^0)/BR(\eta\to\gamma\gamma)$.   The  radiative
width could  be determined  by measuring cross  section of  the $\eta$
meson  two photon  production  using {\it  e.g.}   Primakov effect  or
$e^{\pm}e^-\to e^{\pm}e^-\eta$  process.  The knowledge  of the Dalitz
plot  distributions for  the  $\eta\to3\pi$ decays  will in  principle
contribute  to all  measurements  involving these  final states.   For
example $\Gamma_{\gamma\gamma}$ was  recently extracted from the cross
section of the two  photon production $e^+e^-\to e^+e^-\eta$ where the
$\eta$   meson  was   tagged  by   the   $\eta\to\pi^0\pi^0\pi^0$  and
$\eta\to\pi^+\pi^-\pi^0$ decay modes \cite{Babusci:2012ik}.

The calculations  of the proportionality factors could  be carried out
in the low  energy effective field theory of  the strong interactions,
Chiral Perturbation  Theory (ChPT).  The process was  calculated up to
next-to-next-leading  order  (NNLO) \cite{Bell:1996mi,  Cronin:1967jq,
  Gasser:1984pr, Bijnens:2007pr}.  The  ChPT leading order (LO) result
together  with the  measured  value of  the $\eta\to  \pi^+\pi^-\pi^0$
decay  width of  $300\pm12$ eV  \cite{PDG14} leads  to $Q$  equal 15.6
(Tab.~\ref{table:QandR}).   The next-to-leading order  (NLO) gives
the $Q$ value 28\% larger where  half of the increase comes
from    $\pi\pi$    re-scattering    between   final    state    pions
\cite{Roiesnel:1980gd,Gasser:1984pr}. Finally the NNLO order increases
the value  by an  additional 14\%.  The  values of $Q$  extracted from
various analyses are summarized in Tab.~\ref{table:QandR}.

\begin{table}
\begin{center}
\begin{tabular}{l|l}
	\hline
\textbf{\em Calculations:}  & \textbf{\em Q} \\
	\hline \hline
\small LO \cite{Bijnens:2007pr} & 15.6 \\
\small NLO \cite{Bijnens:2007pr} & 20.1 \\
\small NNLO \cite{Bijnens:2007pr} & 22.9 \\
\hline
\small dispersive  \cite{Anisovich:1996tx} & $22.7(8)$ \\
\small dispersive \cite{Kambor:1995yc} & $22.4(9)$  \\
\small dispersive (PLM) \cite{Kampf:2011wr} & $23.1(7)$\\
\hline
\small Lattice QCD av. \cite{Aoki:2013ldr} & $22.6(7)(6)$\\
	\hline 

	\end{tabular}
\end{center}
	\label{table:QandR}
	\caption{Values   of  $Q$   obtained   from  the
            $\eta\rightarrow 3  \pi$ decay. In addition  a lattice QCD
            estimate is shown for comparison.}
\end{table}

The  reliability of  the calculations  leading to  the proportionality
factor could  be tested by comparing the  experimental and theoretical
Dalitz plots for both the  neutral and charged modes.  Such comparison
constitutes  a sensitive  test of  the convergence  of the  SU(3) ChPT
expansion.  
For the neutral decay mode, where the Dalitz plot density is described
by a single parameter up to quadratic terms, the experiments provide a
consistent,      precise     value     \cite{Alde:1984wj,Abele:1998yi,
  Achasov:2001xi,   Tippens:2001fm,  Bashkanov:2007aa,  Adolph:2008vn,
  Prakhov:2008ff,   Unverzagt:2008ny,   Ambrosinod:2010mj}.   However,
reproduction of  this value has turned  out to be a  challenge for the
ChPT calculations.  For the $\eta\to\pi^+\pi^-\pi^0$ decay mode, where
the  are more  parameters to  describe Dalitz  plot density,  there is
basically    only    one    modern,   high    statistics    experiment
\cite{Ambrosino:2008ht}.

The amplitudes  for the $\eta\to3\pi$ decays could  be also determined
using unitarity and  analyticity and the $\pi \pi$  phase shifts up to
some  subtraction  constants.   These  subtraction  constants  can  be
determined  by  matching  to  the  results of  the  ChPT  calculations
\cite{Kambor:1995yc,Anisovich:1996tx}  and thus  improving convergence
of the ChPT expansion.  Alternatively the subtraction constants can be
obtained  directly   from  fits   to  the  experimental   Dalitz  plot
distributions using only the  most reliable constraints from ChPT.  In
recent years two such  data driven dispersive approaches have emerged:
from Bern-Lund-Valencia  (BLV) group \cite{Colangelo:2009db}  and from
Prague-Lund-Marseille    (PLM)   group    \cite{Kampf:2011wr}.    Both
approaches rely in  large extent on the experimental  Dalitz plot data
from and promise a precise determination of $Q$.

Another aspects of the $\eta\to
3\pi$ decay such as isospin violation effects in low-energy $\pi\pi$ scattering
are addressed by   Non-Relativistic
Effective Field Theory  (NREFT), developed first for low-energy $\pi\pi$ scattering and decay  
$ K  \rightarrow 3\pi$
\cite{Colangelo:2006va}  decays and  subsequently  was applied  to $\eta\rightarrow
3\pi$  decays \cite{Gullstrom:2008sy,Schneider:2010hs}. 
A more model dependent analysis providing  uniform treatment of
all three pseudoscalar 
$\eta$ and $\eta'$ decay modes, including $\eta\rightarrow 3\pi$ was
pursued in  Ref.~\cite{Borasoy:2005du}.   

The  Dalitz plot  for $\eta\to  \pi^+\pi^-\pi^0$  is expressed
using normalized variables $X$ and $Y$:
\begin{equation}
X=\sqrt{3}\frac{T_+ - T_-}{Q_{\eta}}\label{eqn:xy};\ \ \
Y=\frac{3 T_0}{Q_{\eta}}-1,
\end{equation}
where \textit{$T_{+}$,  $T_{-}$, $T_{0}$} are kinetic  energies of the
charged and neutral pions in the $\eta$ meson  rest frame. $Q_{\eta}$ is the excess energy
for the decay:
\begin{equation}
Q_{\eta}=T_{+}+T_{-}+T_{0}\label{eqn:qeta}
\end{equation}
or equivalently $Q_{\eta}=m_{\eta}-2m_{\pm}-m_{0}$ where  $m_{\pm}$ and $m_{0}$
are the masses of the charged and neutral pions.
A polynomial  parametrization is often  used to represent  the squared
amplitude for the decay:
\begin{equation}
|\mathcal{A}(X,Y)|^{2}    \propto   \rho(X,    Y)=    N\left(1   +aY    +
bY^2+cX+dX^2+eXY+fY^3+gX^2Y+hX^3\right),\label{eqn:DPpolform}
\end{equation}
where $\rho(X,  Y)$ is  the Dalitz plot  density, $N$ is  a normalization
factor  and  $a,  b,  ...,  g,h$ are  {\it Dalitz  plot  parameters}.
The  terms with  odd powers  of the $X$  variable, like
$c$,  $e$ and  $h$, should be zero  as they  imply charge  conjugation
violation in strong or  electromagnetic interactions.  The Dalitz plot
parameters from  various theoretical predictions  and from experiments
are given in Tab.~\ref{table:DalPlotparametersTheory}.

\begin{table}
\begin{center}
\begin{tabular}{l| l l l l l}
	\hline\hline
\textbf{\em Calculations}   & \boldmath {$-a$} & \textbf{\em b} &\textbf{\em d} & \textbf{\em f} & \textbf{\em g}  \\
	\hline 
 LO \cite{Bijnens:2007pr} &  $1.039$ & $0.27$ & $0.000$ & 0.000 & $-$\\
 NLO \cite{Bijnens:2007pr}  & $1.371$ & $0.452$ &  $0.053$ & $0.027$ &  $-$ \\ 
 NNLO \cite{Bijnens:2007pr}  & 1.271(75) &  0.394(102) &  0.055(57) &  0.025(160) &  $-$  \\
 dispersive \cite{Kambor:1995yc} &  1.16 &  0.26 &  0.10 &  $-$ & $-$ \\ 
 tree disp \cite{Bijnens:2002qy} & 1.10 &  0.31 &   0.001 &  $-$ & $-$\\ 
 abs disp \cite{Bijnens:2002qy}  &  1.21 &  0.33 &   0.04 &  $-$ &  $-$\\ 
 NREFT \cite{Schneider:2010hs}  &  1.213(14) &  0.308(23) &  0.050(3) &  0.083(19) &  $-0.039(2)$\\
 BSE \cite{Borasoy:2005du}  &   1.054(25) &  0.185(15) &   0.079(26) &  0.064(12) &   $-$\\
	\hline 
\textbf{\em Experiment}  & \boldmath {$-a$} &\textbf{\em b} & \textbf{\em d} &\textbf{\em f} &\textbf{\em g} \\
	\hline 

 Gormley \cite{Gormley:1970qz} &  $1.17(2)$ &  $0.21(3)$ &  $0.06(4)$ &  $-$ &  $-$\\
 Layter $\textit{et al}$ \cite{Layter:1973ti} &  1.080(14) &  0.03(3) &   0.05(3) &  $-$ &  $-$ \\
 CBarrel-98 \cite{Abele:1998yj} &  1.22(7) &  0.22(11) &   $0.06$ (fixed) &  $-$ &  $-$ \\ 
 KLOE \cite{Ambrosino:2008ht} & $1.090(5)(^{+19}_{-8})$ & $0.124(6)(10)$ &  $0.057(6)(^{+7}_{-16})$ & $0.14(1)(2)$ & $ \sim 0$\\
	\hline \hline

	\end{tabular}
\end{center}
	\caption{Dalitz  plot parameters from  theoretical predictions
          and experimental  results for  $\eta \to \pi^+ \pi^- \pi^0$. Results at LO, NLO  and NNLO ChPT
          are taken from  \cite{Bijnens:2007pr}. The values inside the
          parentheses  denote the quoted  uncertainties.  For  the  KLOE data
          both statistical and systematic uncertainties are given.
	\label{table:DalPlotparametersTheory}}
\end{table}
The best precision in the experimental Dalitz plot parameter values is
achieved  in the recent  KLOE \cite{Ambrosino:2008ht}  experiment from
the analysis of 1.34$\cdot10^6$ $\eta \to \pi^+\pi^-\pi^0$ decays. The
description of the  KLOE data requires inclusion of  a cubic term (the
$f$   parameter).   The   quadratic  term   $b$  disagrees   with  the
experimental   results   from   the  seventies   \cite{Gormley:1970qz,
  Layter:1973ti},  while it  agrees  with the  Crystal Barrel  results
\cite{Abele:1998yj}  within uncertainties.  A  comparison of  the KLOE
result to the theoretical  predictions shows disagreement  for
both the $a$ and $b$ parameter  values when taking into account the  
combined uncertainties of the
experimental and  theoretical predictions. The  discrepancies are more
than  five standard  deviations for  the  NNLO parameter  $a$ and  $b$
values.  Also, model independent relations between neutral and charged
Dalitz plot parameters show tensions \cite{Schneider:2010hs}.

A solid experimental data base  for the Dalitz plot distributions is a
must for the  further more detailed investigations.  The  next goal is
to reach  a comparable experimental  status for the charged  $\eta \to
\pi^+\pi^-\pi^0$    channel   as   for    the   neutral    $\eta   \to
\pi^0\pi^0\pi^0$.  Therefore, several new high statistics measurements
of the charged channel are required.

Here  we present  a first  step to  match the  KLOE precision  with an
independent measurement of the  $\eta \to \pi^+\pi^-\pi^0$ Dalitz plot
parameters.

\section{Experiment}
\subsection{The WASA detector}

The   presented  results   are   obtained  with   the  WASA   detector
\cite{Bargholtz:2008ze,Adam:2004ch}, in  an internal target experiment
at  the  Cooler  Synchrotron  COSY storage  ring  \cite{Maier:1997zj},
Forschungszentrum J\"ulich,  Germany.  The COSY  proton beam interacts
with  an  internal  target  consisting  of  small  pellets  of  frozen
deuterium  (diameter $\sim$  35 $\mu$m).   The $\eta$  mesons  for the
$\eta\rightarrow   3\pi$  decay  studies   were  produced   using  the
$pd\to\mbox{}^{3}\mbox{He}\eta$ reaction at a proton kinetic energy of
1 GeV, corresponding to a  center-of-mass excess energy of 60 MeV. The
cross  section  of  the  reaction  is 0.40(3)~$\mu$b  at  this  energy
\cite{Bilger:2002aw,Rausmann:2009dn}.

The  WASA  detector consists  of a Central  Detector  (CD) and  a Forward
Detector      (FD),      covering      scattering     angles
of 20$^{\circ}$--169$^{\circ}$ and 3$^{\circ}$--18$^{\circ}$ respectively
in  combination with  an almost  full azimuthal  angle  coverage.  The
Central Detector is  used to detect and measure  the decay products of
the  mesons.  A  straw  cylindrical  chamber (MDC)  is  placed in  a
magnetic field,  provided by a superconducting  solenoid, for momentum
determination of charged particles.  The central value of the magnetic
field  was   0.85  T  during  the   experiment.   The  electromagnetic
calorimeter   consists   of   1012   CsI(Na)  crystals   read-out   by
photomultipliers.  A plastic scintillator barrel is placed between the
MDC  and the  solenoid allowing  particle identification  and accurate
timing for charged tracks.   The Forward Detector consists of thirteen
layers of plastic scintillators  providing energy and time information
and a straw tube tracker for  precise track reconstruction.

At the  trigger level events  with at least  one track in  the forward
detector and  with a  high  energy deposit  in  thin plastic  scintillator
layers were accepted.  The condition is  effective for selection
of $^3$He  ions and provides an  unbiased data sample  of $\eta$ meson
decays.     The      proton    beam   energy was chosen so  the
$\mbox{}^{3}\mbox{He}$    produced     in    the    $pd    \rightarrow
\mbox{}^{3}\mbox{He}   \eta$  reaction   stop  in   the   first  thick
scintillator layer of the Forward Detector.

\begin{figure}[htb] 
 \centering \includegraphics[width=0.49\textwidth]{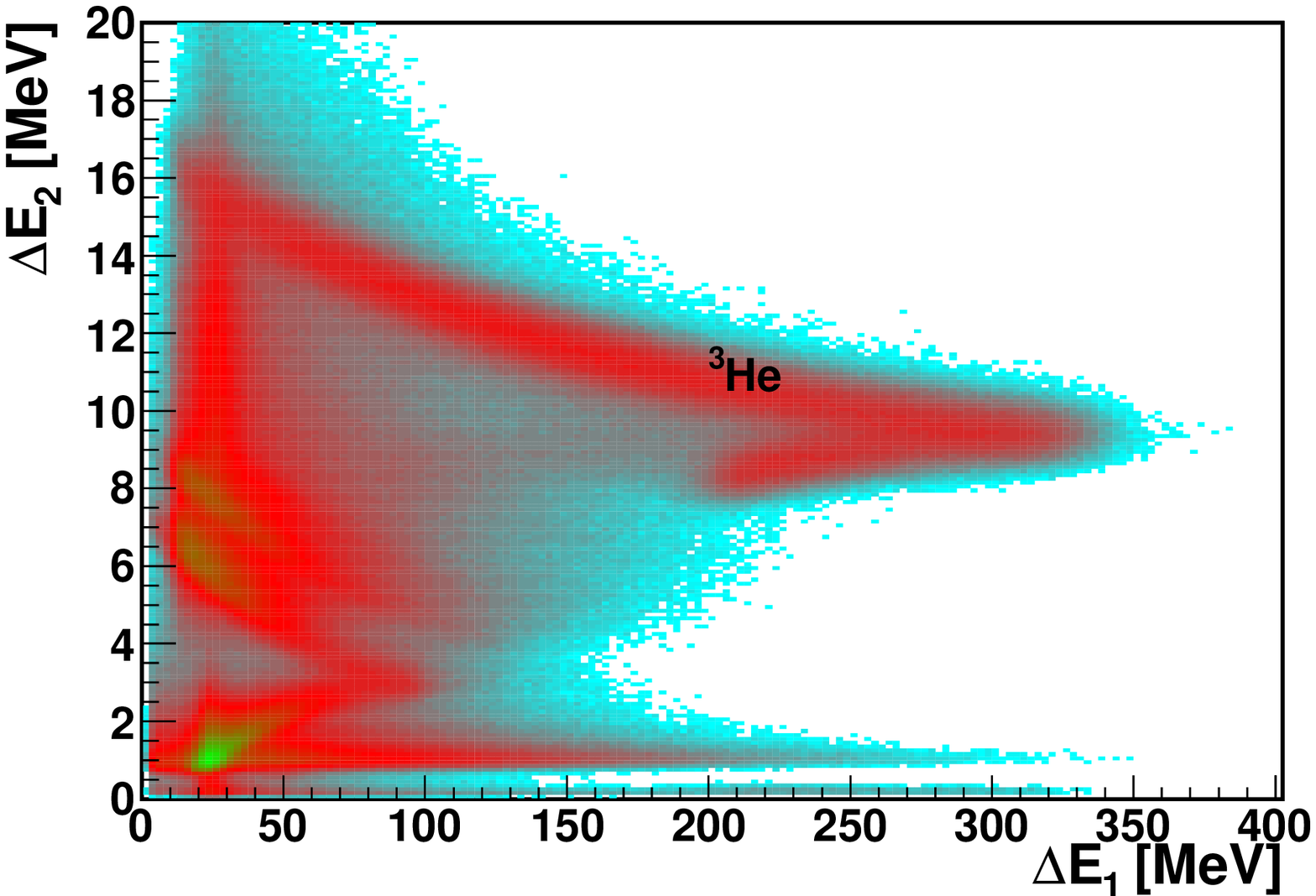}\includegraphics[width=0.49\textwidth]{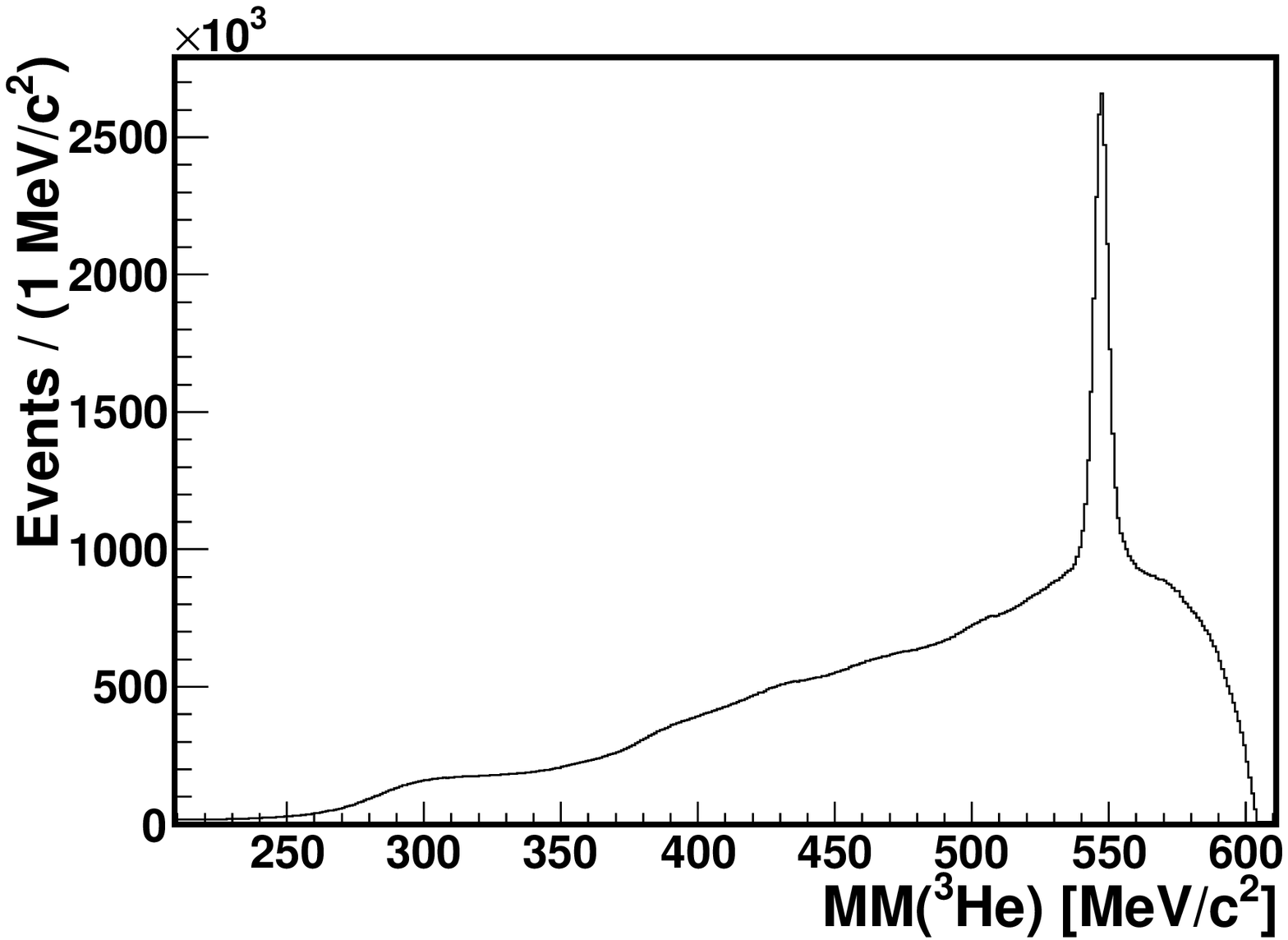} 
 \caption{(left) Correlation of energy  deposits between two Forward Detector
   plastic  detector layers: the first  thick  layer (11  cm), $\Delta
   E_1$, and a preceding thin (0.5 cm) layer, $\Delta E_2$. 
 \label{FVC} 
(right) MM($\mbox{}^{3}\mbox{He}$)    for   all   events    with   a
   $\mbox{}^{3}\mbox{He}$  detected in  the FD.   There  are about
   $1.2\cdot   10^7$    events   in   the    peak   corresponding   to
   the $pd\to\mbox{}^{3}\mbox{He}\eta$ reaction.}
 \label{MMHe} 
\end{figure}

The correlation  plot $\Delta  E- \Delta  E$ from a  thin layer and  the first
thick layer  of the  FD is shown  in Fig.~\ref{FVC}(left). The  (upper) band
corresponding to the $\mbox{}^{3}\mbox{He}$ ion is well separated from
the bands for  other particles and allows a  clear identification of
$\mbox{}^{3}\mbox{He}$.   The   $\mbox{}^{3}\mbox{He}$   from  the
reaction of interest has  kinetic energies ranging between 220~MeV and
460~MeV   and   scattering   angles   ranging  from   $0^{\circ}$   to
$10^{\circ}$.

The missing  mass calculated  from the reconstructed  $^3$He momentum,
MM($\mbox{}^{3}\mbox{He}$), is  shown in Fig.~\ref{MMHe}(right).   The $\eta$
peak has a width of  6.2~MeV$/c^2$ (FWHM) and contains about $1.2\cdot
10^7$  events.   The luminosity  during  the  run  was kept  in  range
$(1-5)\times 10^{31}$ cm$^{-2}$s$^{-1}$.

\subsection{Simulation}
\label{sec:simulation}
The  measurement   of  the  production reaction $pd\to\mbox{}^{3}\mbox{He}\eta$
 is  simulated  by using the experimental angular distribution
from   \cite{Bilger:2002aw,Rausmann:2009dn}.    The   decay  $\eta\rightarrow
\pi^+\pi^-\pi^0$ (BR=22.92(28)\% \cite{PDG14}) was simulated at
the final stage using the central values of the extracted experimental
Dalitz plot parameters.
The  main physics  background processes include  the $\eta\to
\pi^+ \pi^-  \gamma$ (BR=4.22(8)\% \cite{PDG14}) decay  and the direct
two  and three  pion  production reactions: $pd\rightarrow  \mbox{}^{3}\mbox{He}
\pi^+ \pi^-$, $pd\rightarrow  \mbox{}^{3}\mbox{He} \pi^+ \pi^- \pi^0$.
For  the $\eta\rightarrow  \pi^+  \pi^- \gamma$  we  used the  results
reported  in \cite{Adlarson:2011xb,Babusci:2012ft}.  All  other $\eta$
decay  channels contribute  marginally  to the  final  result and  may
therefore  be   neglected.   The  direct   3$\pi$  production  channel
simulated  with uniform  phase space  distributions were  modified to
reproduce  our final MM($\mbox{}^{3}\mbox{He}$)  distribution as  extracted
from Fig.~\ref{KFMMH2}.

The chance coincidental events for the 16 most prominent $pd$ reaction
channels (total cross section 80  mb) and the effect of energy pile-up
in  the   different  detector  elements  are  also   included  in  the
simulation.  Their  relative strengths  of the different  channels are
assumed using  the Fermi statistical model.  For  the quasi-free break
up reactions  the relative momentum  between the np-pair  is simulated
using the deuteron wave function  is used while for all other channels
uniform phase space is assumed.

The accelerator and the target pellet beam overlap region is 3.8~mm in
the horizontal  and 5~mm in  the vertical direction.   The interaction
point  distribution can  have  tails in  the  $z$-direction since  the
accelerator  beam can  also  interact  with a  small  fraction of  the
surrounding rest gas or divergent  pellets.  The shape of the tails is
based on  the $z$-vertex  distribution deduced from  experimental data
with $\mbox{}^{3}\mbox{He}$ production.

\subsection{Event selection}

The signature  of an event, in addition  to the $\mbox{}^{3}\mbox{He}$
ion  reconstructed in FD,   is  at least  two tracks  from charged
particles in the MDC and at  least two clusters in the calorimeter not
associated  with the tracks.   The polar  angles of  charged particles
detected  in  the MDC  are  larger  than  $30^{\circ}$ and  less  than
$150^{\circ}$.  The  time window  in the CD  with respect to  the time
signal  of  the  $\mbox{}^{3}\mbox{He}$  is  6.2~ns  for  the  charged
particle  tracks and  30~ns for  neutral particle  hit.   All possible
combinations of tracks are retained  for kinematic fitting even if the
number of tracks  in the event is greater than  the expected number of
final state particles.

\begin{figure}[htbp] 
 \centering 
 \includegraphics[width=0.49\textwidth]{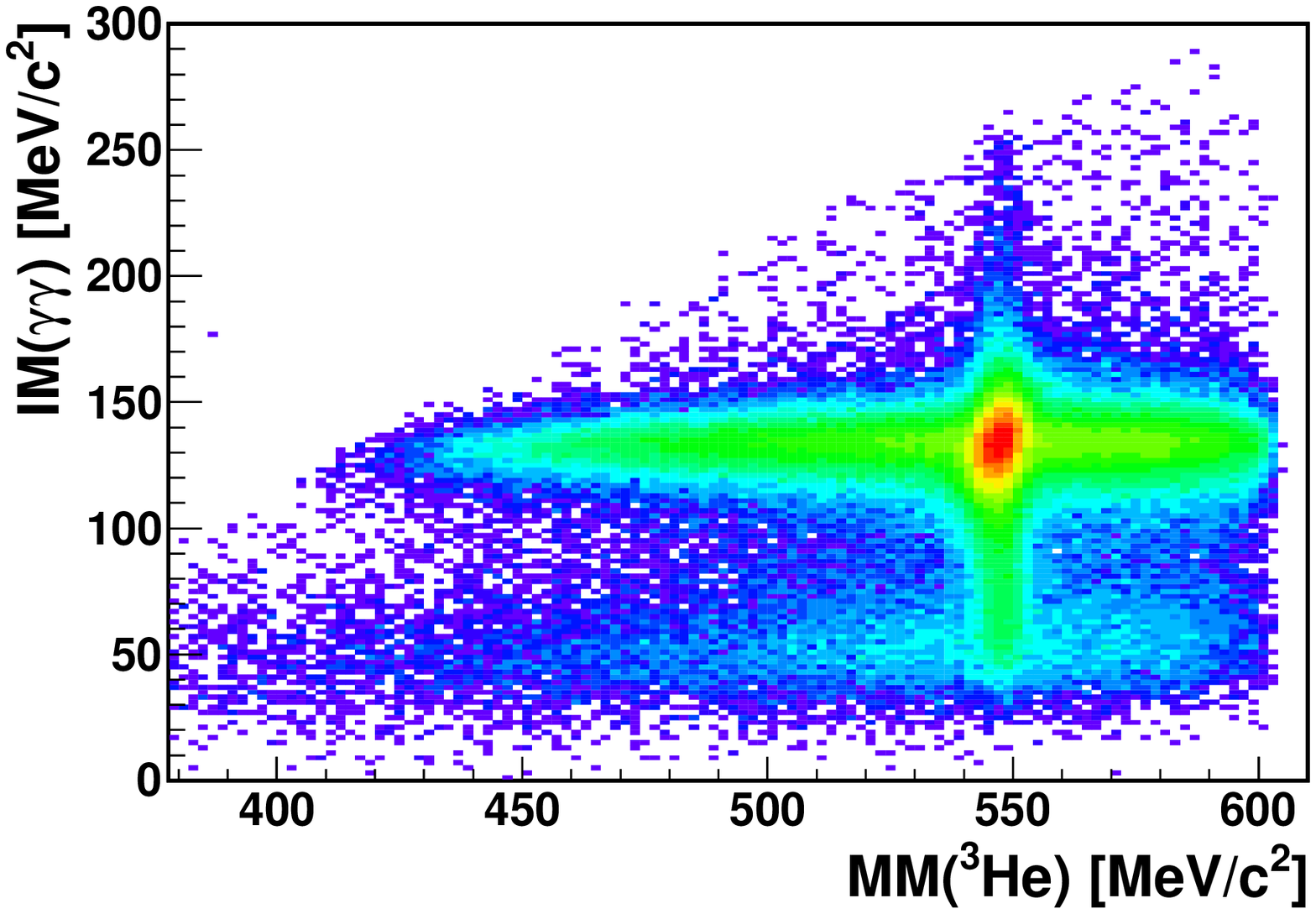}\includegraphics[width=0.49\textwidth]{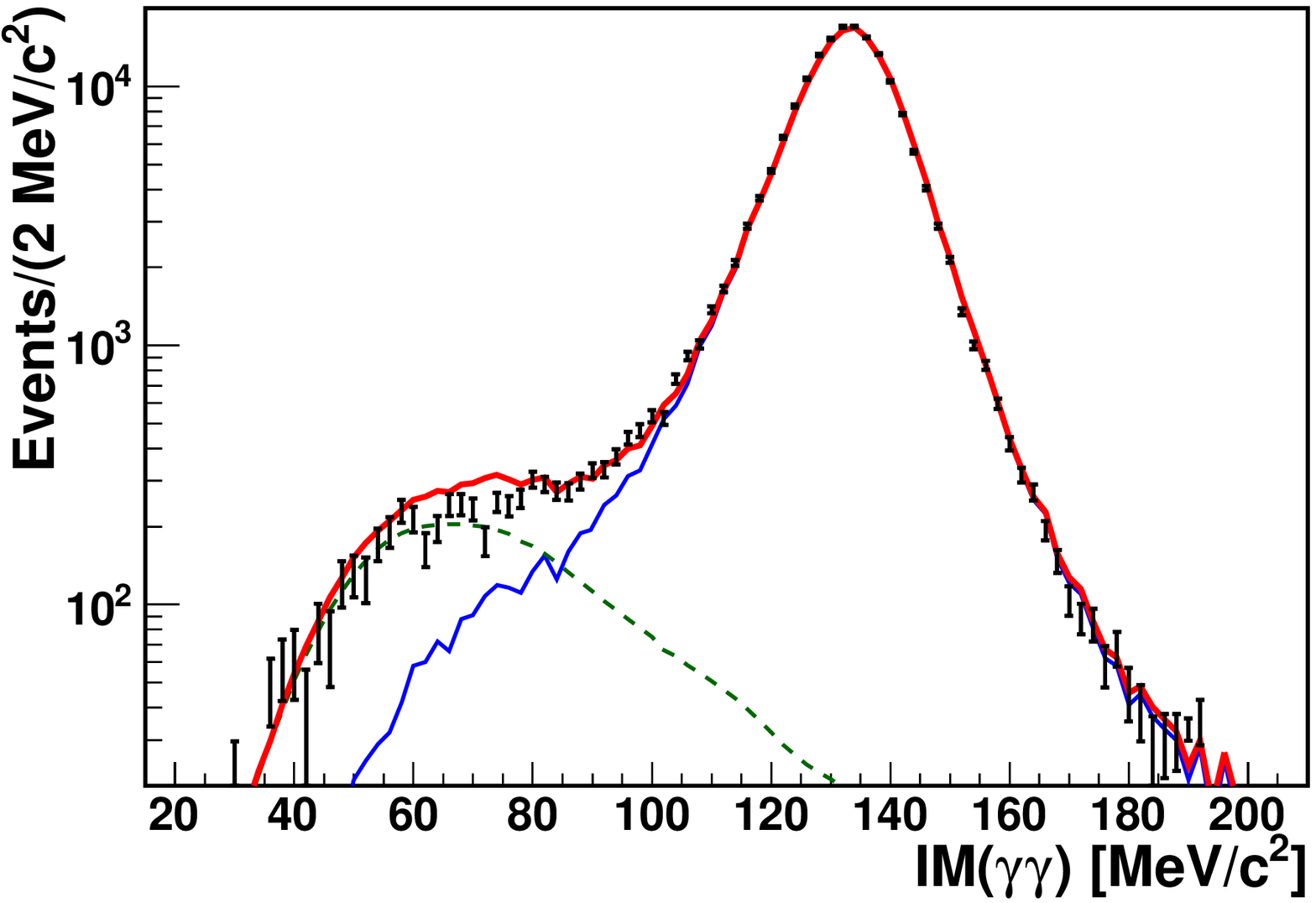}

 \caption{ (left)    Correlation     between     MM$(^3\textrm{He})$     and
   IM$(\gamma\gamma)$  for variables adjusted by  a kinematic
   fit.  \label{fig:IMGGMM}
(right) Comparison of  the experimental and  simulated contributions
   of the  $\eta$ events as  the function of  IM$(\gamma\gamma)$.  The
   extracted  number of  the  events in  the  $\eta$ peak  for each  2
   MeV/c$^2$  IM$(\gamma\gamma)$  slice   is  well  described  by  the
   simulation (thick solid red  line) including the $\eta\to\pi^+\pi^-\pi^0$
   (solid  blue  line)  and $\eta\to\pi^+\pi^-\gamma$  decays  (dashed
   green line). }
 \label{fig:KFMMGG} 
\end{figure}

The point  of closest approach of  the two charged  particle tracks of
the CD  should be within 7~cm from  the center of the  pellet and COSY
beams overlap region. A kinematic fit with the
\begin{equation}
pd\rightarrow \mbox{}^{3}\mbox{He}
\pi^+  \pi^- \gamma  \gamma\label{eqn:kfit}
\end{equation}
 reaction  hypothesis is  applied  and the
combination with the lowest $\chi^2$  value is selected.  A cut on the
$\chi^2$ probability  is made at  1\%.  In the remaining  analysis the
variable values adjusted by the fit are used.   The correlation between the
fitted MM$(^3\textrm{He})$ and the  invariant mass of the two photons,
IM$(\gamma\gamma)$, is shown in Fig.~\ref{fig:IMGGMM}(left).

Fig.~\ref{fig:KFMMGG}(right)  shows   the  extracted  yield   of  the  $pd\to
\mbox{}^{3}\mbox{He}\eta$ events as  a function of IM$(\gamma\gamma)$.
The  distribution  was obtained  by  creating  2 MeV/c$^2$  horizontal
slices of  the scatter  plot in Fig.~\ref{fig:IMGGMM}(left)  and determining
the peak content of each  one.  The resulting distribution agrees well
with     simulations    of     the     $\eta\to\pi^+\pi^-\pi^0$    and
$\eta\to\pi^+\pi^-\gamma$ decays.   The relative normalization between
the two  decays is fixed by  their branching ratios. For  the final data
sample only events with IM$(\gamma\gamma)>100$ MeV/c$^2$ are selected.

\begin{figure}[htbp] 
 \centering 
 \includegraphics[width=0.7\textwidth]{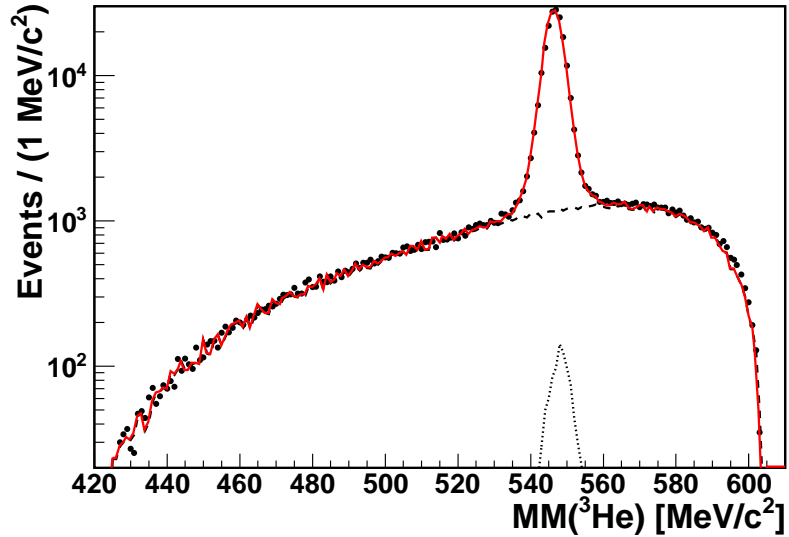} 
 \caption[]{The   distribution  of   MM$(\mbox{}^{3}\mbox{He})$  using
   variables adjusted by  the kinematic fit for the  final data sample
   (dots), agree  well with the  sum of the Monte  Carlo distributions
   for the signal and the backgrounds (red solid line). Separately are
   shown contributions  from $\eta\rightarrow\pi^+\pi^-\gamma$ (dotted
   line) and from the direct $3\pi$ production (dashed line). }
 \label{KFMMH2} 
\end{figure}

The data  sample used  in this analysis  consists of  $1.74\cdot 10^5$
$\eta$ candidates.  The comparison  of the simulated  and experimental
distributions    of    MM$(\mbox{}^{3}\mbox{He})$    is    shown    in
Fig.~\ref{KFMMH2}.  The dominating  background comes from direct three
pion production.   The contributions from two pion  production and the
$\eta\to\pi^+\pi^-\gamma$ decay are less than 1$\%$.
\section{Results}

The  variables $X$  and  $Y$ are  calculated from  Eqn.~(\ref{eqn:xy})
using the  kinetic energies of  the charged pions after  the kinematic
fitting  boosted to  the  rest frame  of the  $\pi^+\pi^-\gamma\gamma$
system.  For  the variables  after the kinematic  fit of  the reaction
(\ref{eqn:kfit}) one  has $\mu\equiv \mbox{IM}(\pi^+\pi^-\gamma\gamma)
=  \mbox{MM}(^{3}\mbox{He})$.  However,  $\mu$ is  not  constrained to
equal  $m_\eta$  and  IM$(\gamma\gamma)$  not  constrained  to  $m_0$.
Therefore,  the  kinetic  energy  of  the  neutral  pion,  $T_{0}$,  is
determined in the following way:
\begin{equation}
T_{0}=\mu-T_{+}-T_{-}-2m_{\pm}-\mbox{IM}(\gamma\gamma),
\end{equation}
and for calculating $Q_{\eta}$ we use Eqn.~(\ref{eqn:qeta}).

\begin{figure}[htbp] 
 \centering 
 \includegraphics[width=0.6\textwidth]{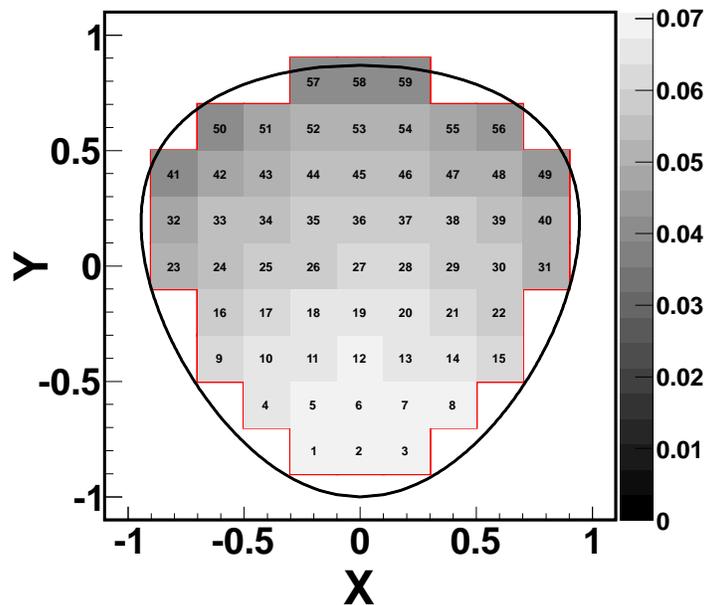} 
 \caption{ Position and  numbering of the Dalitz plot  bins used for
   the  analysis.   The  acceptance for  the  $\eta\to\pi^+\pi^-\pi^0$
   decay is also indicated by the gray scale.  }
 \label{fig:acc} 
\end{figure}

The selected  Dalitz plot bin width  in $X$ and  $Y$ ($\Delta X=\Delta
Y=0.2$) is in our case limited by the statistics needed for background
subtraction and reliable systematical crosschecks.  The uncertainty of
the $X$ and $Y$ measurement is well within the experimental resolution
(FWHM  of approximately 0.10  for both  $\Delta X$  and $\Delta  Y$ in
average).   The $X,Y$  region $[-1.1,1.1]\times[-1.1,1.1]$  is divided
into $11\times 11$  bins.  The border bins with  less than 90\% Dalitz
plot area inside  the kinematic boundaries are excluded  leading to 59
bins used in the analysis.  Definition and the numbering scheme of the
bins is given in Fig.~\ref{fig:acc}.

The Dalitz plot for  the $\eta\to\pi^+\pi^-\pi^0$ decay is obtained by
dividing  the  reconstructed  $X$  and  $Y$ variables  into  bins  and
determining  the signal  content in  each bin  from  the corresponding
$\mu$ distribution.  The signal content  in each bin is estimated by a
least    squared   fit    of    the   simulated    data   of    $pd\to
\mbox{}^{3}\mbox{He}\eta$  and the  $pd\to  \mbox{}^{3}\mbox{He} \pi^+
\pi^-  \pi^0$  continuum  background  reaction.   The  matrix  element
squared of the background reaction  is assumed to be a linear function
of $\mu$:
\begin{equation}
  F_i(\mu)=N_S^is_i(\mu)+N_B^i\left(1+\alpha_i \mu\right)b_i(\mu),
\label{eqn:fitmm}
\end{equation}
where $i$ is  the Dalitz plot bin number,  $N_S^i$ is the normalization
factor  for  the  simulated $pd\to  \mbox{}^{3}\mbox{He}\eta$  signal,
$s_i(\mu)$.  $N_B^i, b_i(\mu)$ have the corresponding meaning with respect
to    the    flat   phase    space    simulation    of   the    $pd\to
\mbox{}^{3}\mbox{He}\pi^+ \pi^- \pi^0$ reaction.  $N_S^i$, $N_B^i$
and $\alpha_i$ are free parameters in the fit.

\begin{figure}[htbp] 
 \centering 
\includegraphics[width=1.0\textwidth]{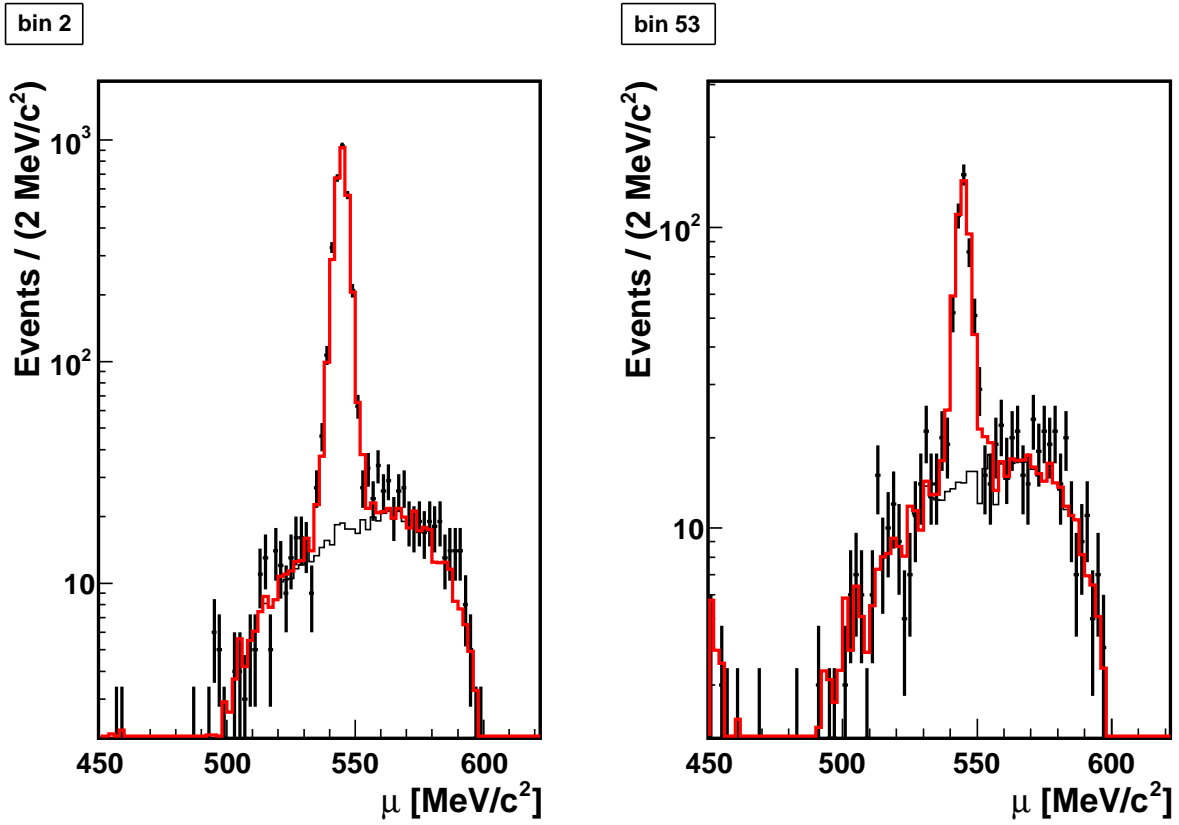} 
 \caption{Two examples of the  fits to the $dN/d\mu$ distributions for
   a large (left) and a low statistics Dalitz plot bin (right).
The red thick line is the fitted function Eq.~\ref{eqn:fitmm} while
the thin line represents the continuous background contribution.}
 \label{polfitex} 
\end{figure}

Two examples of  the fits are shown in  Fig.~\ref{polfitex}; one for a
Dalitz  plot bin  with larger  statistics (bin \#2, centered  at $X=0$,
$Y=-0.8$) and one  for a bin with lower  statistics, (bin \#53, centered
at $X=0$, $Y=0.6$).

Finally the simulated background from $\eta\to\pi^+\pi^-\gamma$ events
is subtracted  from $N_S^i$.  This  contribution is small  compared to
the  statistical uncertainties.   The extracted  number of  $\eta\to 3
\pi$ events is  corrected for acceptance.  It
was checked that  the use of bin by  bin acceptance correction ({\it i.e.}
diagonal   smearing  matrix)  does   not  introduce   any  significant
systematic effect.

\begin{figure}[ht] 
 \centering 
 \includegraphics[width=0.9\textwidth]{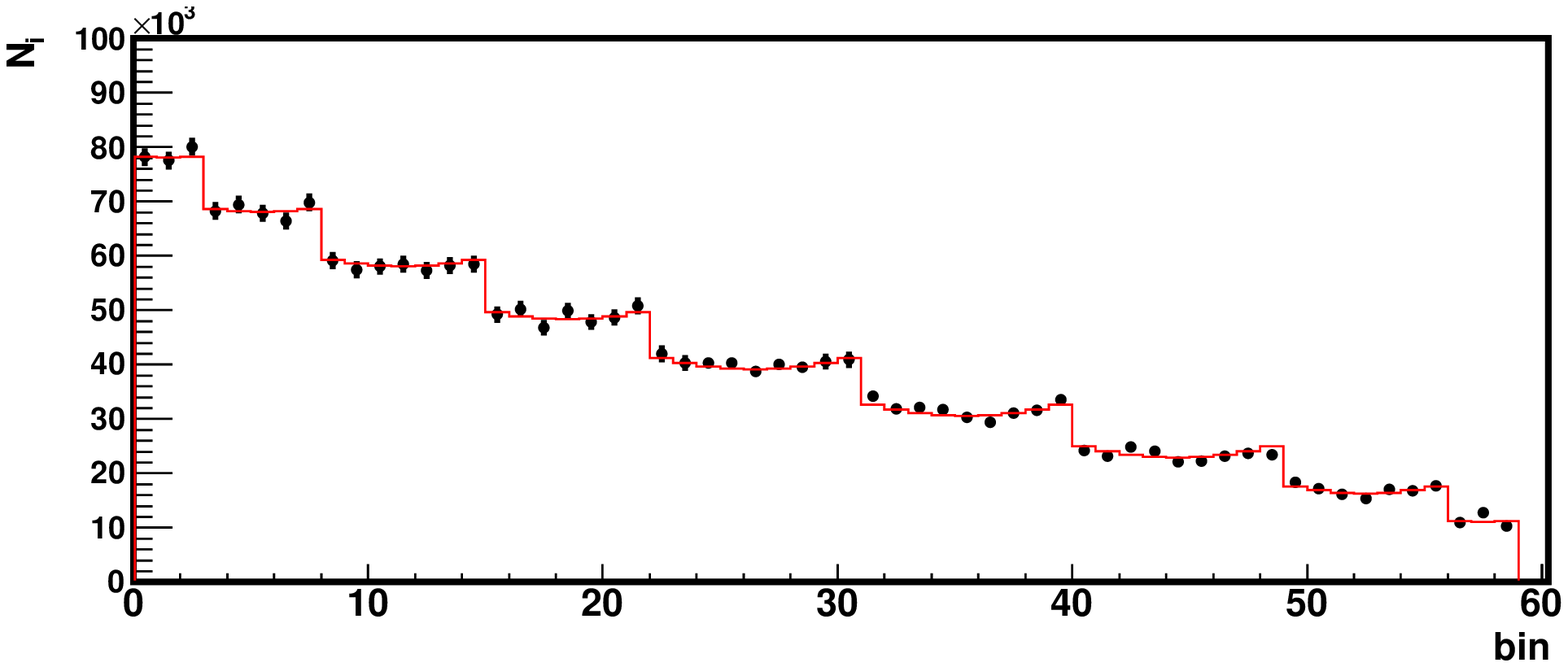} 
 \includegraphics[width=0.9\textwidth]{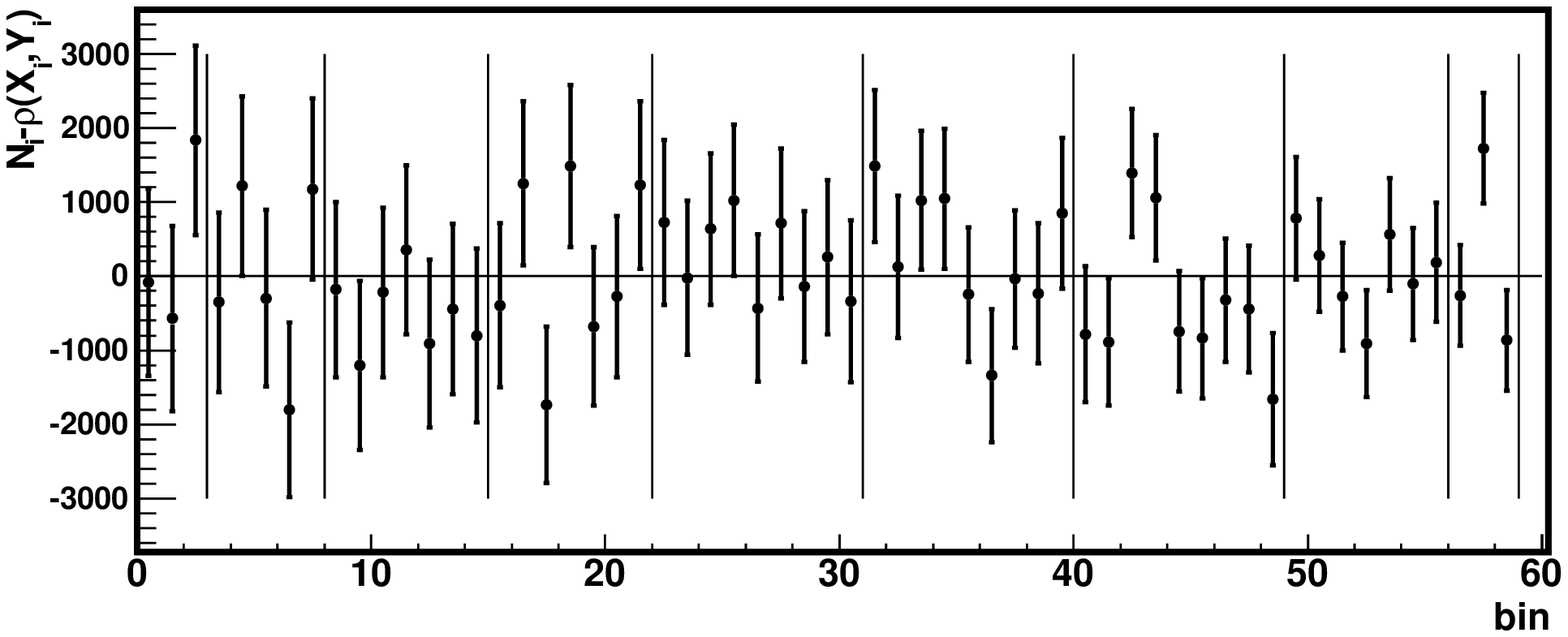} 
 \caption{The upper  panel shows acceptance corrected  Dalitz plot bin
   contents  with statistical uncertainties  (black points  with error
   bars) compared to the fitted function $\rho(X, Y)$ (red line) for each
   bin.  The lower panel shows the corresponding residuals.}
 \label{DPslices120521} \label{fig:DPstd1d}
\end{figure}

The acceptance  values, indicated in  Fig.~\ref{fig:acc}, are obtained
from a MC sample of $5\cdot$10$^7$ $\eta\to\pi^+\pi^-\pi^0$ events and
varies between 4\% and 7\%.  It  is larger when $T_{0}$ is small ({\it i.e.}
lower  $Y$-values), but  also when  the  kinetic energies  of the  two
charged   pions  are   similar  ({\it i.e.}    for  $X$   close   to  zero).
Fig.~\ref{fig:DPstd1d}  shows  the   acceptance  corrected  number  of
$\eta\rightarrow\pi^+\pi^-\pi^0$ events as function of the Dalitz plot
bin number.

The Dalitz plot parameters are  obtained with the least square fitting
procedure which minimizes
\begin{equation}
\chi^2=\sum^{59}_{i=1}\left(\frac{N_i-\rho(X_i, Y_i)}{\Delta N_i}\right)^2.
\end{equation}
$N_i$  and $\Delta  N_i$  denote the  acceptance  corrected number  of
events  and their  statistical uncertainty  for the  Dalitz  plot bins
$i=1,..,59$).    The    function   $\rho(X_i,   Y_i)$,    defined   in
Eqn.~(\ref{eqn:DPpolform}), is evaluated at  the center of each Dalitz
plot  bin:  $X_i$ and  $Y_i$.   In  our  case the  systematic  effects
introduced by this procedure are negligible as it was checked using MC
data  sample.  The  overall normalization  factor $N$  is also  a free
parameter in the fit.

The obtained  Dalitz plot  parameters together with  their statistical
uncertainties  are  presented  in Tab.~\ref{table:eta3pistandard}  for
different assumptions  about the Dalitz plot  parameters together with
the fit  $\chi^2$ and number of  degrees of freedom  ({\it dof}).  The
$c$ and $e$  parameters are fixed to $0$ in the  fits.  In addition we
have performed fits including  these parameters.  The result gives $c$
and  $e$ consistent  with zero:  $c=-0.007(9)$ and  $e=-0.020(23)$ and
does not  affect other  parameters.  For the  case when all  $a$, $b$,
$c$, $d$, $e$ and $f$  parameters are fit one obtains $\chi^2/dof=46.6
/ 52$.  The  correlation matrix between the fitted  parameters for the
standard result obtained is shown in Tab.~\ref{table:corrmatrix}.
\begin{table} [htb] 
\begin{center}
\begin{tabular}{l| l l l l l l c }
	\hline \hline
	& $a$ & $b$  & $d$ & $f$ & $\chi^2/dof$ \\
	\hline 
 4 parameters ({\it std})&$-1.144(18)$ &  0.219(19)  &  0.086(18)&   0.115(37) &  49.4 / 54\\
 3 parameters  &  $-1.101(11)$ &  0.234(19)  &   0.078(18)&   0 (fix)&  58.8 / 55\\
 2 parameters &  $-1.075(9)$ &  0.201(17)  &   0 (fix) &   0 (fix) &  78.3 / 56\\
	\hline\hline 
	\end{tabular}
\end{center}
	\caption[]{Fit   results  for   different  sets of  Dalitz  plot
          parameters.  The normalization  factor, $N$, is omitted from
          the  table. A  number  followed by  '(fix)'  means that  the
          corresponding parameter  was fixed to this  number.
	\label{table:eta3pistandard}}
\end{table}

\begin{table} [htb] 
\begin{center}
\begin{tabular}{l|r r r  }

   & \hspace*{0.8cm}\textbf{\em a} &  \hspace*{0.8cm}\textbf{\em b} & \hspace*{0.8cm}\textbf{\em d} \\ 
\hline
 \textbf{\em b}\hspace*{0.3cm} & $-0.24$ &   &       \\ 
\textbf{\em d}\hspace*{0.3cm}  & $-0.45$ & $0.36$ &     \\ 
\textbf{\em f}\hspace*{0.3cm} & $-0.79$ & $-0.25$ &  $0.14$     \\ 
	\end{tabular}
\end{center}
	
	\caption{Correlation matrix for the Dalitz plot parameters.\label{table:corrmatrix}}
\end{table}
Tab.~\ref{table:corrmatrix}  shows a  strong  anti-correlation between
the parameter $a$ and $f$ which is also reflected in the uncertainties
of the  parameter $a$.  The  bins of the  Dalitz plot are  compared in
Fig.~\ref{DPslices120521}  to  the   parametrization  with  four  free
parameters  $a,b,d,f$  where  the  remaining  ones  are  set  to  zero
(parametrization       labeled       as       {\it       std}       in
Tab.~\ref{table:eta3pistandard}).

\section{Systematic uncertainties}

The systematic  uncertainties of  the obtained Dalitz  plot parameters
are  investigated  by including  variations  due  to  know sources  of
uncertainties in the  MC generated data and by  changing the selection
criteria to find the remaining effects.  In particular the consistency
of  extraction of  the Dalitz  plot  distribution and  fitting of  the
Dalitz plot parameters  were tested using MC generated  data ten times
larger than  in the experiment.  The input  parameters were reproduced
without introducing any  systematical deviation within the statistical
uncertainties.

One of the most important sources of systematical uncertainties is the
direct background subtraction procedure. This uncertainty is estimated
by comparing  a fit with the  signal region excluded from  the fit and
the signal  term $N_S^is_i(\mu)$ in  Eqn.~(\ref{eqn:fitmm}) is omitted
and the background  is subtracted directly from the  data ([Test 1] in
Tab.~\ref{table:eta3pisysteffectsold}).


To investigate  further possible systematical effects  the data sample
has  been  divided  into  sets   of  high  and  low  luminosity.   The
$pd\rightarrow X$ cross section is  $\sim$ 80~mb, which amounts to few
background  reactions produced  per $\mu  s$.  The  largest  effect is
connected to the calorimeter signals  since the decay times are of the
order of $\mu  s$.  The Dalitz plot parameter  values obtained for the
low   and  high   luminosity  sample   are  shown   by  [Test   2]  in
Tab.~\ref{table:eta3pisysteffectsold}.


Two different  accelerator beam modes  were used during the  beam time
and they cover roughly   equal time  of  data taking.   In  the  first half,  a
constant  beam energy  during the  accelerator cycle  was assured  by a
fixed radio frequency (RF).  In  the second half, a coasting beam
 with RF switched  off swept the target leading to a slight decrease of
the beam energy during a cycle (from 1000.0 MeV to 993.5~MeV).  In the
experimental analysis this energy decrease  is taken into account.  However,
in  the simulations  the acceptance  has been calculated for a  beam
kinetic energy fixed at 1~GeV.  The comparison of the two cases ([Test
  3] in Tab.~\ref{table:eta3pisysteffectsold}) shows the largest deviation
for  the  $b$ parameter  ($\approx$  2$\sigma$).   To investigate  the
source of the  effect we have calculated the  acceptances also for the
lowest beam  energy in the RF  off mode (993.5~MeV)  and concluded that the
change is too small to explain the observed deviation.

The effect  of the uncertainty of the  implemented detector resolution
in the detector simulations is  tested by increasing the kinematic fit
probability     from     0.01      to     0.1     ([Test     4]     in
Tab.~\ref{table:eta3pisysteffectsold}).   The  difference between  the
parameter values  are not significant  and are therefore  neglected in
the final systematical uncertainty.

\begin{table} [htb] 
\begin{center}
\begin{tabular}{l| c c  c  c | c }
	\hline\hline
	& {\boldmath $-a$} &  \textbf{\em b} & \textbf{\em d} & \textbf{\em f} & $\chi^2$ / $d.o.f.$\\
	\hline
 Standard result &  1.144(18) &  0.219(19)  &  0.086(18)&   0.115(37) &  49.4 / 54\\

[Test 1] Background fit &  1.126(18) &  0.230(19)  & 0.094(18)&   0.111(37) &  60.5 / 54\\

[Test 2] Low luminosity &  1.130(24)&  {0.216(26)}  &  0.059(24)&   0.104(50) &  50.5 / 54\\ 

[Test 2] High luminosity &  {1.164(25)}&  {0.219(28)}  &  {0.106(26)}&   {0.152(52)} &  54.9 / 54\\ 

[Test 3] RF &  1.127(26)&  {0.177(28)}  &  {0.085(27)} &   {0.140(55)} &  56.1 / 54\\ 

[Test 3] No RF &  1.139(23)&  {0.252(26)}  &  0.076(24) &   {0.069(49)} &  49.6 / 54\\

[Test 4]  PDF$>0.1$ &  {1.146(22)} &  0.224(24)  &  {0.075(22)}&   0.117(46) &  48.0 / 54\\
\hline

	\end{tabular}
\end{center}
	
	\caption{ Dalitz plot parameters  extracted for
   different   tests  for   systematic
          effects. Description in the text.\label{table:eta3pisysteffectsold}}
\end{table}




The only  significant changes are seen  for the $b$  parameter for the
two accelerator  operation modes  and for $d$  for the  two luminosity
cases.  We use metodology of Ref.~\cite{Barlow:2002yb} and express the
final result for the Dalitz plot parameters in the following way:
\begin{eqnarray*}
	-a&=&1.144 \pm 0.018(stat)\\
	 b&=&0.219 \pm 0.019(stat)\pm 0.037(syst) \\
	 d&=&0.086 \pm 0.018(stat)\pm 0.018(syst) \\
	 f&=&0.115 \pm 0.037(stat).
\end{eqnarray*}

In addition we give the values  for the C violating parameters $c$ and
$e$:
\begin{eqnarray*}
 	c&=& -0.007\pm 0.009(stat)\\
	e&=& -0.020\pm 0.023(stat)\pm 0.029(syst).
\end{eqnarray*}

The results  are dominated by statistical  uncertainties and therefore
the   provided   table  with   acceptance   corrected  bin   contents,
Tab.~\ref{table:result},  could be used  directly for  comparison with
theoretical models.

\section{Discussion of results}

Parameters $a$, $b$  and $d$ significantly deviate from  zero. The $d$
parameter     is    3.4$\sigma$     above    zero.     From
Tab.~\ref{table:eta3pistandard}  it  is seen  that  $\chi^2$ per  {\it
  dof}  is only slightly worse if parameter $f$ is set to zero in
the fit.  The significance of allowing $f\ne 0$ in our data is 3.1$\sigma$. 
  However, the $a$  and $f$
parameters        are       strongly        anti-correlated       (see
Tab.~\ref{table:corrmatrix})  and excluding $f$  from the  fit affects
also the $a$  value.  The data  do not require  higher order terms  in the
polynomial expansion such as $g\cdot X^2 Y$ and $h\cdot X^3$.

Here we list  deviations from the Dalitz plot  parameters obtained by
the KLOE   collaboration  \cite{Ambrosino:2008ht}   together   with  their
significance (statistical and systematic uncertainties are added in squares):
\begin{eqnarray*}
 -\Delta a&=&+0.054(23)\ \ (+2.3\sigma)\\
  \Delta b&=&+0.095(44)\ \ (+2.2\sigma)\\
  \Delta d&=&+0.029(28)\ \ (+1.0\sigma)\\
  \Delta f&=&-0.025(43)\ \ (-0.6\sigma).
\end{eqnarray*}

Our results  are  generally consistent  with KLOE,  however
there is  some tension for $a$  and $b$ parameters.   Our data confirm
the   discrepancies   between   theoretical   calculations   and   the
experimental values  from the KLOE experiment.   The provided experimental
data points of the individual  Dalitz plot bins will allow independent
analyses using NREFT or dispersive methods.

The presented  results are based on  a first part  of the WASA-at-COSY
data  from  the  $pd\to^3\textrm{He}\eta$  reaction.   More  data  are
available    from    WASA-at-COSY    also    from  the  $pp\to    pp\eta$
reaction. Together  with expected  results from other  experiments the
goal  of a  precise determination  of  the $\eta\to\pi^+\pi^-\pi^0$ Dalitz
plot parameters might soon be reached. 
\begin{table}[htb]
\begin{center}
\begin{tabular}{ r@{\hspace{10pt}}  c  @{\hspace{20pt}}| r @{\hspace{10pt}}  c   @{\hspace{20pt}}| r @{\hspace{10pt}} c  @{\hspace{20pt}}| r @{\hspace{10pt}} c   } \hline\hline
 Bin \# & Content & Bin\# &  Content  & Bin \# &  Content  & Bin \# &  Content \\
 \hline
1 & 2.020 $\pm$ 0.033 & 16 & 1.271 $\pm$ 0.029 & 31 & 1.058 $\pm$ 0.028 & 46 & 0.573 $\pm$ 0.021  \\
2 & 2.004 $\pm$ 0.032 & 17 & 1.296 $\pm$ 0.029 & 32 & 0.883 $\pm$ 0.027 & 47 & 0.597 $\pm$ 0.022  \\
3 & 2.069 $\pm$ 0.033 & 18 & 1.209 $\pm$ 0.027 & 33 & 0.824 $\pm$ 0.025 & 48 & 0.611 $\pm$ 0.022  \\
4 & 1.764 $\pm$ 0.031 & 19 & 1.289 $\pm$ 0.028 & 34 & 0.830 $\pm$ 0.024 & 49 & 0.604 $\pm$ 0.023  \\
5 & 1.794 $\pm$ 0.031 & 20 & 1.236 $\pm$ 0.028 & 35 & 0.820 $\pm$ 0.024 & 50 & 0.473 $\pm$ 0.021  \\
6 & 1.752 $\pm$ 0.031 & 21 & 1.257 $\pm$ 0.028 & 36 & 0.783 $\pm$ 0.024 & 51 & 0.443 $\pm$ 0.020  \\
7 & 1.716 $\pm$ 0.031 & 22 & 1.313 $\pm$ 0.029 & 37 & 0.758 $\pm$ 0.023 & 52 & 0.418 $\pm$ 0.019  \\
8 & 1.804 $\pm$ 0.032 & 23 & 1.085 $\pm$ 0.029 & 38 & 0.802 $\pm$ 0.024 & 53 & 0.398 $\pm$ 0.019  \\
9 & 1.528 $\pm$ 0.031 & 24 & 1.042 $\pm$ 0.027 & 39 & 0.815 $\pm$ 0.025 & 54 & 0.440 $\pm$ 0.020  \\
10 & 1.484 $\pm$ 0.029 & 25 & 1.041 $\pm$ 0.026 & 40 & 0.867 $\pm$ 0.026 & 55 & 0.433 $\pm$ 0.020  \\
11 & 1.499 $\pm$ 0.030 & 26 & 1.041 $\pm$ 0.026 & 41 & 0.626 $\pm$ 0.024 & 56 & 0.458 $\pm$ 0.021  \\
12 & 1.511 $\pm$ 0.030 & 27 & 1.000 $\pm$ 0.026 & 42 & 0.600 $\pm$ 0.022 & 57 & 0.283 $\pm$ 0.018  \\
13 & 1.481 $\pm$ 0.029 & 28 & 1.033 $\pm$ 0.026 & 43 & 0.641 $\pm$ 0.022 & 58 & 0.331 $\pm$ 0.019  \\
14 & 1.504 $\pm$ 0.030 & 29 & 1.021 $\pm$ 0.026 & 44 & 0.622 $\pm$ 0.022 & 59 & 0.268 $\pm$ 0.018  \\
15 & 1.512 $\pm$ 0.030 & 30 & 1.049 $\pm$ 0.027 & 45 & 0.572 $\pm$ 0.021 &  &   \\ \hline\hline
\end{tabular}
\end{center}
\caption[]{ \label{table:result}Acceptance corrected
 Dalitz plot distribution. The bin numbering is given in Fig.~\ref{fig:acc}.
The bin contents are normalized to the bin centered at $X=0,Y=0$ (bin \#27).}
\end{table}

\begin{acknowledgments}

This work  was supported in  part by the EU  Integrated Infrastructure
Initiative    HadronPhysics     Project    under    contract    number
RII3-CT-2004-506078;  by   the  European  Commission   under  the  7th
Framework Programme  through the 'Research  Infrastructures' action of
the  'Capacities' Programme,  Call:  FP7-INFRASTRUCTURES-2008-1, Grant
Agreement N.   227431; by the  Polish National Science  Centre through
the     Grants    No.      86/2/N-DFG/07/2011/0    0320/B/H03/2011/40,
2011/01/B/ST2/00431,   2011/03/B/ST2/01847,   0312/B/H03/2011/40   and
Foundation for Polish Science.   We gratefully acknowledge the support
given by the  Swedish Research Council, the Knut  and Alice Wallenberg
Foundation, and the Forschungszentrum  J\"ulich FFE Funding Program of
the J\"ulich Center for Hadron Physics.

This work is based on the PhD thesis of Patrik Adlarson supported by
Uddeholms Forskarstipendium.
\end{acknowledgments}

\bibliography{pub}
\end{document}